\documentclass[reprint, amsmath,amssymb,aps]{revtex4-2}
\usepackage{graphicx}
\usepackage{dcolumn}
\usepackage{bm}
\usepackage{hyperref}
\hypersetup{
    colorlinks=true,
    linkcolor=blue,
    filecolor=blue,      
    urlcolor=blue,
    citecolor=blue
    }
\usepackage{amsmath,amssymb,amsfonts,bm}
\usepackage{ulem}
\usepackage[dvipsnames]{xcolor}

\begin{document}

\title{Nuclear geometry driven symmetry plane correlations \\ in OO and Ne--Ne collisions at the Large Hadron Collider}

\author{Suraj Prasad$^{1}$}\email[]{Suraj.Prasad@cern.ch}
\author{Raghunath Sahoo$^{2}$}\email[Corresponding Author: ]{Raghunath.Sahoo@cern.ch}
\affiliation{$^1$HUN-REN Wigner Research Centre for Physics, 29-33 Konkoly-Thege Miklós Str., H-1121 Budapest, Hungary}
\affiliation{$^2$Department of Physics, Indian Institute of Technology Indore, Simrol, Indore 453552, India}

\begin{abstract}
Symmetry-plane correlations (SPCs) are key observables sensitive to the medium's transport properties and are driven by participant-plane correlations (PPCs) in the nuclear overlap region. This study explores the possibility of nuclear-geometry-driven SPCs in Oxygen--Oxygen (OO) and Neon--Neon (Ne--Ne) collisions at $\sqrt{s_{\rm NN}}=5.36$ TeV using nuclear geometry simulations based on Nuclear Lattice Effective Field Theory (NLEFT) and Projected Generator Coordinate Method (PGCM) configurations. We investigate $\langle \cos[4(\psi_2 - \psi_4)]\rangle_{\rm GE}$ and $\langle \cos[6(\psi_3 - \psi_6)]\rangle_{\rm GE}$ in OO and Ne--Ne collisions at $\sqrt{s_{\rm NN}}=5.36$ TeV using the A Multi-Phase Transport (AMPT) model. We find that Ne--Ne collisions exhibit larger $\langle \cos[4(\psi_2 - \psi_4)]\rangle_{\rm GE}$ values than OO collisions, whereas $\langle \cos[6(\psi_3 - \psi_6)]\rangle_{\rm GE}$ is larger in OO than in Ne--Ne collisions. This behavior indicates a strongly deformed shape of the $^{20}$Ne nucleus and a tetrahedral structure of the $^{16}$O nucleus. We also explore SPCs for events with tip-tip and body-body collision configurations, which further support these findings.
\end{abstract}
\date{\today}
\maketitle 

\section{Introduction}
\label{intro}
The world's most powerful particle colliders, such as the Large Hadron Collider (LHC) at the European Organization for Nuclear Research (CERN) and the Relativistic Heavy-Ion Collider (RHIC) at Brookhaven National Laboratory (BNL), collide heavy ions to produce and study a deconfined state of strongly interacting matter known as the quark-gluon plasma (QGP). Due to the transient nature of this exotic state, direct observation is not possible. Therefore, experiments rely heavily on different signatures carried by the produced hadrons until they reach the detectors, from which the formation of such a medium can be inferred. One of the most important properties of this strongly interacting matter is its collective behavior, which is usually studied by anisotropic flow coefficients~\cite{Heinz:2013th, Ollitrault:1992bk}. The anisotropic flow coefficients appear as the coefficients in the Fourier expansion of the azimuthal particle distribution, given as follows~\cite{Voloshin:1994mz}.

\begin{equation}
\frac{dN}{d\phi}\propto 1+\sum_{n=1}^{\infty}v_{n}\cos[n(\phi-\psi_{n})]
\end{equation}

Here, $\phi$ denotes the azimuthal angle, and $\psi_{n}$ is the $n$\textsuperscript{th}-order symmetry-plane angle. The coefficient $v_{n}$ represents the $n$\textsuperscript{th}-order anisotropic flow, with $v_2$ corresponding to elliptic flow, $v_3$ to triangular flow, and so on.
The symmetry plane angles and anisotropic flow coefficients can be expressed together using multi-particle correlations as follows~\cite{Bhalerao:2011yg}:
\begin{align}
\label{eq:corrln_def}
        &v_{n_1}^{a_1}v_{n_2}^{a_2}...v_{n_k}^{a_k}e^{i(a_1n_1\psi_{n_1}+a_2n_2\psi_{n_2}+...+a_kn_k\psi_{n_k})} \nonumber\\&= \langle e^{i(n_1\phi_1+n_2\phi_2+...+n_l\phi_l)}\rangle
\end{align}
Here, $a_i$ specifies how many times a given harmonic $n_i$ contributes to the correlation. Equation~\eqref{eq:corrln_def} is essential for constructing multiparticle correlations with reduced contributions from non-flow effects and event-by-event fluctuations through the application of suitable kinematic cuts. Flow coefficients with suppressed non-flow contributions are sensitive not only to the transport properties of the medium but also to the initial-state eccentricities~\cite{Prasad:2025ezg, Giacalone:2021udy, Behera:2023nwj, Prasad:2022zbr, ALICE:2022wpn, Parkkila:2021tqq, Parkkila:2021yha, Prasad:2025yfj}.

Similar to $v_n$, the final-state symmetry-plane angle $\psi_{n}$ is correlated with the corresponding participant-plane angle $\Phi_n$. However, due to the random event-by-event orientation of $\psi_{n}$, its ensemble average, $\langle \psi_{n} \rangle$, approaches zero for sufficiently large event samples~\cite{Wu:2022exl}. In contrast, correlations between different symmetry planes remain finite and are usually studied using symmetry-plane correlations (SPCs)~\cite{E877:1996czs}. These SPCs are primarily driven by the underlying participant-plane correlations (PPCs) and are sensitive to the transport properties of the medium formed in ion collisions. If one assumes factorization between symmetry-plane correlations and the magnitudes of the flow coefficients, the symmetry-plane correlations can be quantified using a Gaussian estimator (GE) of the form~\cite{Bilandzic:2020csw, ALICE:2023wdn}:

\begin{widetext}
\begin{equation}
\label{eqn:SPCGE}
\langle \cos(a_1n_1\psi_{n_1}+a_2n_2\psi_{n_2}+...+a_kn_k\psi_{n_k})\rangle_{\rm GE}
=\sqrt{\frac{\pi}{4}}\frac{\langle{v_{n_1}^{a_1}v_{n_2}^{a_2}...v_{n_k}^{a_k}\cos(a_1n_1\psi_{n_1}+a_2n_2\psi_{n_2}+...+a_kn_k\psi_{n_k})}\rangle}{\sqrt{\left\langle{v_{n_1}^{2a_1}v_{n_2}^{2a_2}...v_{n_k}^{2a_k}}\right\rangle}}
\end{equation}
\end{widetext}

Here, the angular brackets, $\langle \dots\rangle$, denote the event average value.

Unlike anisotropic flow coefficients, which have been measured with statistically significant precision across collision systems ranging from proton-proton (pp) to lead-lead (Pb--Pb)~\cite{ALICE:2022wpn, STAR:2005gfr, Prasad:2025yfj}, symmetry-plane correlations (SPCs) remain relatively unexplored. In the current landscape, where considerable effort is directed toward understanding the origin of apparent collective, fluid-like behavior in small systems at the LHC~\cite{ALICE:2022wpn, Prasad:2025yfj}, SPCs offer an alternative and potentially robust probe of the underlying collective dynamics~\cite{ATLAS:2014ndd, Bilandzic:2020csw, ALICE:2023wdn}. In this context, light-ion collisions such as Oxygen--Oxygen (OO) and Neon--Neon (Ne--Ne) serve as a crucial intermediate testing ground~\cite{Brewer:2021kiv, Citron:2018lsq}. These systems preserve a well-defined initial-state geometry necessary for the development of SPCs while exhibiting system sizes comparable to high-multiplicity proton-lead (p--Pb) collisions. As such, they provide a controlled environment to investigate the onset and evolution of collective phenomena in small to intermediate collision systems.

Additionally, studies in low-energy nuclear physics suggest the possible formation of $\alpha$ ($^{4}$He) clusters in $^{16}$O and $^{20}$Ne nuclei~\cite{GGamowbook, Wheeler:1937zza, Lombardo:2025erk}. In the case of $^{16}$O, the four $\alpha$ clusters are expected to be arranged at the vertices of a regular tetrahedron. In contrast, $^{20}$Ne is often described by a ``pinball"-like configuration, which can be interpreted as a $^{16}$O$+\alpha$ structure. This configuration leads to a pronounced quadrupole deformation in $^{20}$Ne, whereas the tetrahedral arrangement in $^{16}$O gives rise to a significant octupole deformation~\cite{ALICE:2025luc}.

These distinct intrinsic geometries can modify the initial overlap region in nuclear collisions and, consequently, influence final-state observables that are sensitive to the initial geometry. Several theoretical studies have explored the impact of such clustered nuclear structures on final-state flow coefficients in light-ion collisions~\cite{Li:2020vrg,Bozek:2014cva,Broniowski:2013dia,Behera:2023oxe,Ding:2023ibq,Wang:2021ghq,Rybczynski:2017nrx,Svetlichnyi:2023nim,Giacalone:2024ixe,Zhang:2024vkh,R:2024eni, Behera:2023nwj, MenonKavumpadikkalRadhakrishnan:2025apq, Prasad:2024ahm, Behera:2021zhi}. Experimentally, these effects have been investigated at the LHC by ALICE, ATLAS, and CMS~\cite{ALICE:2025luc, ATLAS:2025nnt, CMS:2025tga}. In particular, the measured ratio of $v_2$ in Ne--Ne to OO collisions, when compared with model calculations, points toward a strongly elongated (quadrupole-deformed) shape of the $^{20}$Ne nucleus~\cite{ALICE:2025luc, ATLAS:2025nnt}. Similarly, the relatively smaller ratio of $v_3$ in Ne--Ne to OO collisions is consistent with expectations from a tetrahedral configuration of $^{16}$O nuclei.

Similarly, SPCs are highly sensitive to the transverse overlap geometry in nuclear collisions and are therefore expected to provide strong constraints on the intrinsic structure of the colliding nuclei~\cite{Tripathy:2025npe}. To disentangle the effects of nuclear geometry on SPC observables, this study investigates symmetry-plane correlations in OO and Ne--Ne collisions at $\sqrt{s_{\rm NN}}=5.36$ TeV using the A Multi-Phase Transport (AMPT) model. The initial nuclear configurations are modeled using Nuclear Lattice Effective Field Theory (NLEFT)~\cite{Frosini:2021fjf, Frosini:2021sxj} and the Projected Generator Coordinate Method (PGCM)~\cite{Lahde:2019npb, Lee:2020meg}, both incorporating $\alpha$-cluster parameterizations, and are subsequently implemented as input to the AMPT framework~\cite{Giacalone:2024luz, Loizides:2025ule}. The SPCs are studied using the Gaussian Estimator, shown in Eq.~\eqref{eqn:SPCGE}. The analysis focuses on two specific SPC observables, $\langle \cos[4(\psi_2 - \psi_4)]\rangle_{\rm GE}$ and $\langle \cos[6(\psi_3 - \psi_6)]\rangle_{\rm GE}$, chosen for their sensitivity to the underlying nuclear deformations. In particular, these correlators are expected to probe the quadrupole deformation of the $^{20}$Ne nucleus and the octupole deformation associated with the tetrahedral structure of $^{16}$O.

The rest of the paper is organized as follows. We present the event generation and methodology in Sec.~\ref{sec_method}. The results are shown and discussed in Sec.~\ref{sec_results}. Finally, the paper is summarized in Sec.~\ref{sec_summary}.

\section{Event Generation and Methodology}
\label{sec_method}

\subsection{A multi-phase transport model}
This study is performed using simulated data generated using A Multi-Phase Transport (AMPT) model in its string-melting configuration. AMPT provides a microscopic description of the collision evolution from the initial state to final hadrons~\cite{Zhang:1999bd, Lin:2004en}. The initial particle production is modeled through HIJING~\cite{Wang:1991hta}. The partonic stage is then evolved via Zhang's Parton Cascade (ZPC), which accounts for parton scatterings~\cite{Zhang:1997ej}. Hadronization is implemented through a quark-coalescence mechanism after partonic freeze-out~\cite{Andersson:1983ia, Greco:2003xt}. Finally, the hadronic phase is subsequently treated with the A Relativistic Transport (ART) hadron cascade, including hadron--hadron rescattering~\cite{Li:1995pra,Li:2001xh}.

For the present study, OO and Ne--Ne events are generated at $\sqrt{s_{\rm NN}}=5.36$ TeV. The NLEFT and PGCM nuclear configurations are used as input geometry to quantify structure-driven effects, which are taken from Refs.~\cite{Giacalone:2024luz, Loizides:2025ule}. The AMPT settings are kept same for both the collision systems to isolate genuine nuclear-geometry contributions in the studied SPCs, which is similar to described in Refs.~\cite{Behera:2021zhi,Behera:2023nwj}.

\subsection{Isolating tip-tip and body-body collisions}

In relativistic nuclear collisions, tip-tip and body-body configurations correspond to extremes of the initial geometric overlap of the colliding nuclei. Tip-tip (t-t) collisions are characterized by a small geometric eccentricity but a higher transverse density, leading to larger radial pressure. In contrast, body-body (b-b) collisions exhibit large ellipticity with lower transverse density. To preferentially select these configurations, one can utilize the reduced flow vector $q_2$ and the mean transverse momentum $\langle p_{\rm T} \rangle$ of charged particles, which are probes for eccentricity and transverse radial pressure, respectively.

The second-order reduced flow vector is defined as~\cite{STAR:2002hbo}:
\begin{equation}
q_2 = \frac{1}{\sqrt{M}} \left| \sum_{j=1}^{M} e^{2 i \phi_j} \right|,
\end{equation}
where $M$ is the number of charged particle in the event and $\phi_j$ is the azimuthal angle of particle $j$. 

The mean transverse momentum is defined as follows.
\begin{equation}
\langle p_{\rm T} \rangle = \frac{1}{M} \sum_{i=1}^{M} p_{{\rm T},i}
\end{equation} 

By taking the ratio,
\begin{equation}
R \equiv \frac{q_2}{\langle p_{\rm T} \rangle},
\end{equation}
one constructs an event-shape observable that combines ellipticity and radial flow. In similar collision centrality, events with higher ${q_2}/{\langle p_{\rm T} \rangle}$ correspond to body-body collisions, where the large eccentricity dominates relative to radial push. Similarly, the lower value of ${q_2}/{\langle p_{\rm T} \rangle}$ selects tip-tip collisions, which have smaller eccentricity but stronger radial flow. Thus, sorting events by $q_2 / \langle p_{\rm T} \rangle$ effectively biases the sample towards t-t or b-b orientations.

\section{Results and Discussions}
\label{sec_results}
In relativistic heavy-ion collisions, the measured higher-order anisotropic flow coefficients depend upon corresponding initial eccentricities and receive significant nonlinear contributions from lower-order harmonics~\cite{ATLAS:2015qwl}. These nonlinear mode couplings give rise to correlations between the specific symmetry-plane angles, which can be quantified via the study of SPCs. Therefore, in this study, the choice of the SPCs $\langle \cos[4(\psi_2 - \psi_4)] \rangle_{\rm GE}$ and $\langle \cos[6(\psi_3 - \psi_6)] \rangle_{\rm GE}$ are motivated by their direct sensitivity to such nonlinear coupling mechanisms that connect lower- and higher-order harmonics. Moreover, the strength of these couplings depends not only on the collective dynamics of the system but also on the structure of the collision overlap geometry. In the presence of specific nuclear configurations, such as clustered or deformed nuclei, the eccentricities are affected. This leads to specific correlations between participant plane angles, which in turn affect the coupling between different harmonics and is reflected in the corresponding SPC observables. Consequently, these SPCs provide a sensitive probe of the underlying nuclear geometry, enabling the investigation of geometric features such as quadrupole and octupole deformations in light and heavy-ion collisions.

\begin{figure*}[!ht]
    \centering
    \includegraphics[width=0.8\linewidth]{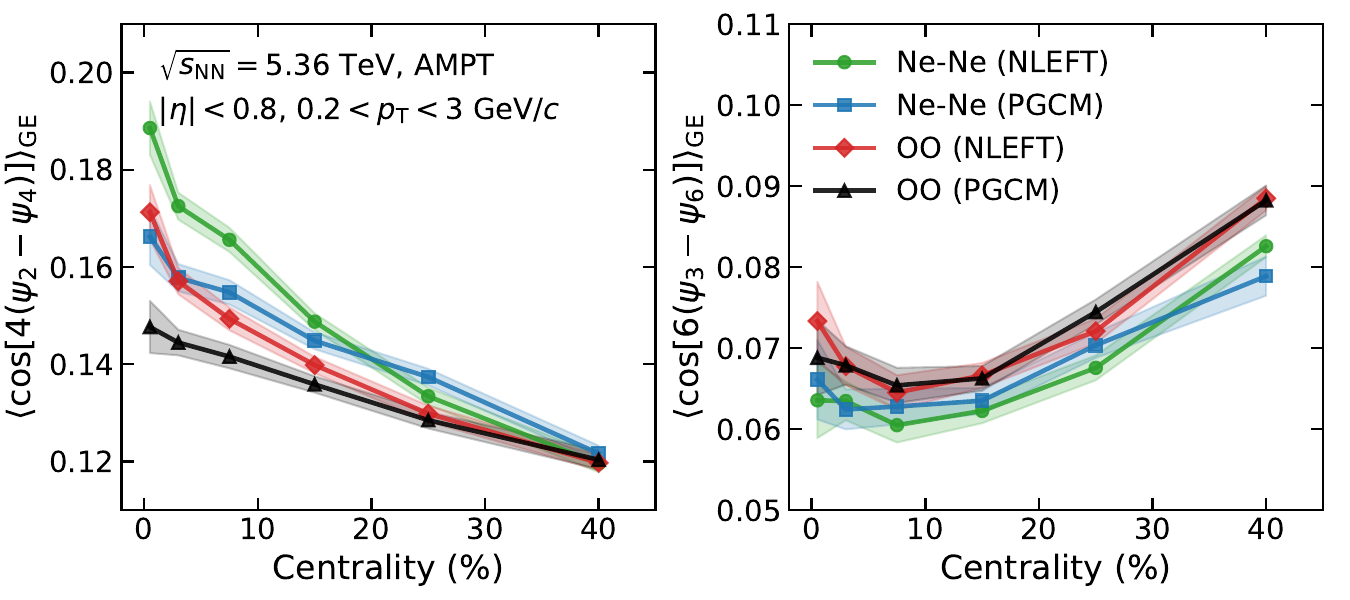}
    \caption{$\langle\cos[4(\psi_2-\psi_4)]\rangle$ (left) and $\langle\cos[6(\psi_3-\psi_6)]\rangle$ (right) as a function of centrality in OO and Ne--Ne collisions at $\sqrt{s_{\rm NN}}=5.36$ TeV using AMPT.}
    \label{fig:1}
\end{figure*}

\begin{figure*}[!ht]
    \centering
    \includegraphics[width=0.7\linewidth]{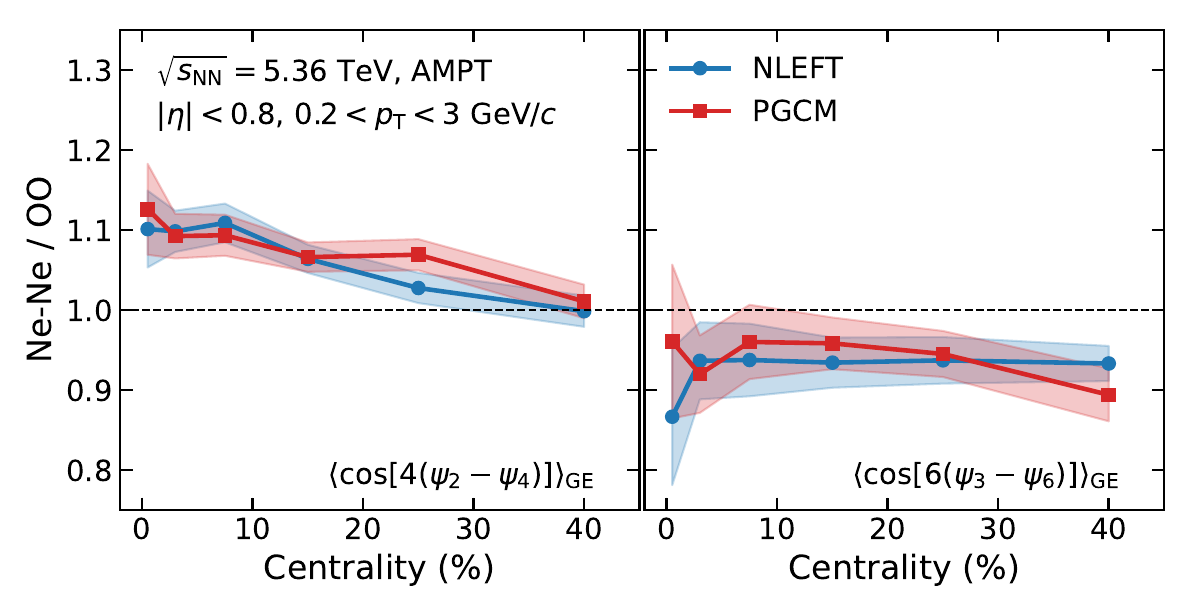}
    \caption{The ratio of SPCs in Ne--Ne to OO collisions for $\langle\cos[4(\psi_2-\psi_4)]\rangle$ (left) and $\langle\cos[6(\psi_3-\psi_6)]\rangle$ (right) as a function of centrality at $\sqrt{s_{\rm NN}}=5.36$ TeV using AMPT.}
    \label{fig:2}
\end{figure*}

To understand these effect of initial nuclear geometry, Fig.~\ref{fig:1} shows the SPCs $\langle \cos[4(\psi_2 - \psi_4)] \rangle_{\rm GE}$ (left) and $\langle \cos[6(\psi_3 - \psi_6)] \rangle_{\rm GE}$ (right) in OO and Ne--Ne collisions at $\sqrt{s_{\rm NN}}=5.36$ TeV using AMPT. In Fig.~\ref{fig:1}, a contrasting centrality dependence is observed for the $\langle \cos[4(\psi_2 - \psi_4)] \rangle_{\rm GE}$ and $\langle \cos[6(\psi_3 - \psi_6)] \rangle_{\rm GE}$, reflecting their different underlying physical origins. The correlator $\langle \cos[4(\psi_2 - \psi_4)] \rangle_{\rm GE}$ decreases towards peripheral collisions, which can be attributed to the reduced contribution of nonlinear mode coupling of fourth and second harmonics in smaller and more dilute systems. In such collisions, the collective response weakens and fluctuation-driven contributions to the fourth harmonics become relatively more abundance, leading to a weaker correlation between $\psi_2$ and $\psi_4$. In contrast, the correlator $\langle \cos[6(\psi_3 - \psi_6)] \rangle_{\rm GE}$ shows a decrease followed by a increasing trend towards peripheral collisions. This behavior reflects the dominance of fluctuation-driven triangular symmetry, where the third harmonic remains significant even in small systems. The sixth-order harmonic receives a substantial nonlinear contribution from the third order harmonics, which induces a strong alignment between $\psi_6$ and $\psi_3$. The opposite centrality trends of these two SPCs thus highlight the interplay between geometry-driven and fluctuation-driven harmonics, with $\langle \cos[4(\psi_2 - \psi_4)] \rangle_{\rm GE}$ primarily sensitive to elliptic geometry, while $\langle \cos[6(\psi_3 - \psi_6)] \rangle_{\rm GE}$ provides a robust probe of fluctuation-induced triangular structure.

Rather than the dependence on collision centrality, the values of SPCs near the central collisions are important as they carry the signatures of nuclear deformations.
In the left panel of Fig.~\ref{fig:1}, one can observe that for a particular nuclear configuration and towards the central collisions, $\langle \cos[4(\psi_2 - \psi_4)] \rangle_{\rm GE}$ for Ne--Ne collisions is higher than that of OO collisions. This is because, the SPC $\langle \cos[4(\psi_2 - \psi_4)] \rangle_{\rm GE}$ probes the coupling between the second- and fourth-order symmetry planes. In particular, the fourth-order flow harmonic receives a dominant nonlinear contribution proportional to the square of the elliptic flow, $v_4 \sim \chi_{4,22}(v_2)^2$, which results in an alignment between $\psi_4$ and $\psi_2$~\cite{Teaney:2013dta, Gardim:2011xv}. Therefore, this observable quantifies the strength of ellipticity-driven coupling and serves as a sensitive probe of the quadrupole component of the initial geometry. In the case of Ne--Ne collisions, where the $^{20}$Ne nucleus exhibits a pronounced quadrupole deformation associated with its clustered (pinball-like) structure, a stronger elliptic anisotropy is expected. This leads to an enhanced nonlinear coupling and consequently larger values of $\langle \cos[4(\psi_2 - \psi_4)] \rangle_{\rm GE}$. In contrast, for OO collisions, where the intrinsic geometry is dominated by tetrahedral clustering with a relatively weaker quadrupole component, the correlation strength is expected to be reduced. From the left panel of Fig.~\ref{fig:2}, it is clear that the effect is stronger in the central collisions, whereas towards the peripheral collisions, the values of in Ne--Ne and OO become closer.

The correlator $\langle \cos[6(\psi_3 - \psi_6)] \rangle_{\rm GE}$, on the other hand, probes the coupling between the third- and sixth-order symmetry planes and is primarily sensitive to triangular symmetry in the system. The sixth-order harmonic receives a nonlinear contribution of the form $v_6 \sim \chi_{6,33}(v_3)^2$, which induces a correlation between $\psi_6$ and $\psi_3$~\cite{Teaney:2013dta, Gardim:2011xv}. As a result, this observable reflects the strength of triangular symmetry and provides access to the octupole component of the initial geometry. In OO collisions, the tetrahedral configuration of the $^{16}$O nucleus naturally gives rise to a significant intrinsic triangular symmetry, leading to a stronger $v_3$ and hence an enhanced $\langle \cos[6(\psi_3 - \psi_6)] \rangle_{\rm GE}$, as seen in the right panel of Fig.~\ref{fig:1}. In contrast, the pinball-like structure of $^{20}$Ne does not favor a similarly strong and coherent triangular deformation, which results in a comparatively weaker correlation. This is reflected in the right panel of Fig.~\ref{fig:2}, where the ratio of SPCs in Ne--Ne to OO collisions are shown.

\begin{figure*}
    \centering
    \includegraphics[width=0.8\linewidth]{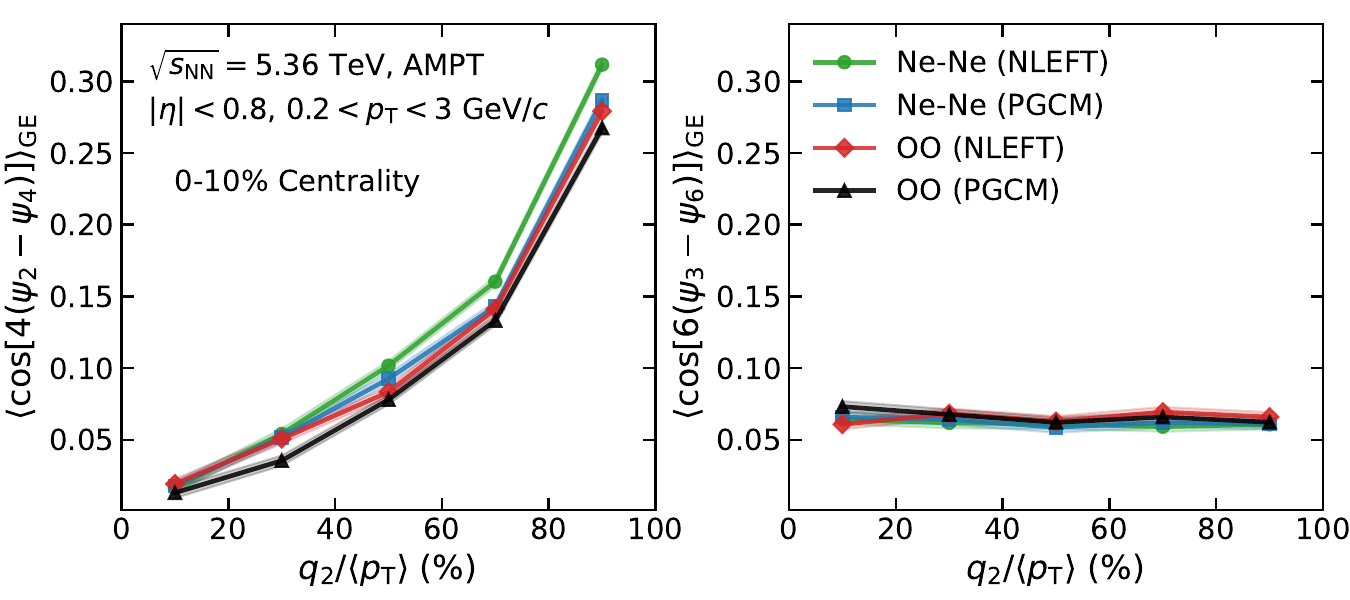}
    \caption{$\langle\cos[4(\psi_2-\psi_4)]\rangle$ (left) and $\langle\cos[6(\psi_3-\psi_6)]\rangle$ (right) as a function of $q_{2}/\langle p_{\rm T}\rangle$ percentile class for 0-10\% central classes in OO and Ne--Ne collisions at $\sqrt{s_{\rm NN}}=5.36$ TeV using AMPT.}
    \label{fig:3}
\end{figure*}

\begin{figure*}
    \centering
    \includegraphics[width=0.7\linewidth]{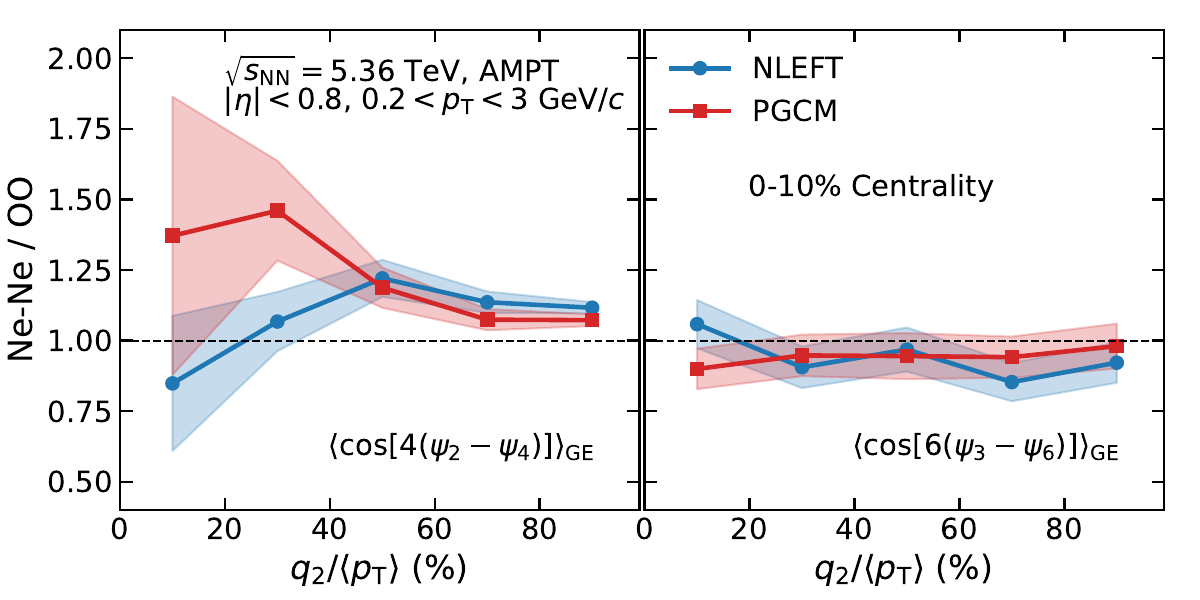}
    \caption{The ratio of SPCs in Ne--Ne to OO collisions for $\langle\cos[4(\psi_2-\psi_4)]\rangle$ (left) and $\langle\cos[6(\psi_3-\psi_6)]\rangle$ (right) as a function of $q_{2}/\langle p_{\rm T}\rangle$ percentile classes for 0-10\% central class at $\sqrt{s_{\rm NN}}=5.36$ TeV using AMPT.}
    \label{fig:4}
\end{figure*}

To further explore the influence of nuclear deformation and cluster structures in light nuclei, we study the SPCs separately for tip-tip and body-body collisions. This classification is particularly relevant for central collisions, where the colliding nuclei interact nearly head-on. In such events, the geometric configuration of the nuclei is imprinted on the collision overlap region and subsequently transmitted to the final-state hadrons through partonic collectivity, allowing the effects of nuclear geometry in different planes to be probed via different SPCs. 

Figure~\ref{fig:3} shows {$\langle\cos[4(\psi_2-\psi_4)]\rangle$ (left) and $\langle\cos[6(\psi_3-\psi_6)]\rangle$ (right) as a function of $q_{2}/\langle p_{\rm T}\rangle$ percentile class for 0-10\% central classes in OO and Ne--Ne collisions at $\sqrt{s_{\rm NN}}=5.36$ TeV using AMPT. Using $q_2 / \langle p_{\rm T} \rangle$ as a proxy to isolate tip-tip and body-body collisions, the correlator $\langle \cos[4(\psi_2 - \psi_4)]\rangle_{\rm GE}$ is observed to increase with $q_2 / \langle p_{\rm T} \rangle$ for both Ne--Ne and OO collisions, reflecting the stronger nonlinear coupling between $\psi_2$ and $\psi_4$ in body-body collisions with larger ellipticity. Tip-tip collisions, in contrast, have smaller geometric eccentricity and weaker alignment, leading to a smaller $\langle \cos[4(\psi_2 - \psi_4)]\rangle_{\rm GE}$. The slightly higher values of $\langle \cos[4(\psi_2 - \psi_4)]\rangle_{\rm GE}$ in Ne--Ne are consistent with the stronger intrinsic quadrupole deformation of $^{20}$Ne, whose pinball-like nuclear structure enhances the ellipticity-driven coupling. Therefore, the $\langle \cos[4(\psi_2 - \psi_4)]\rangle_{\rm GE}$ correlator acts as a sensitive probe of the quadrupole deformation and allows discrimination between tip-tip and body-body orientations.

In contrast, $\langle \cos[6(\psi_3 - \psi_6)]\rangle_{\rm GE}$, as shown in right panel of Fig.~\ref{fig:3}, remains nearly independent of $q_2 / \langle p_{\rm T} \rangle$ and collision system within uncertainties. This independence becomes evident from the right panel of Fig.~\ref{fig:4}. This independence of $\langle \cos[6(\psi_3 - \psi_6)]\rangle_{\rm GE}$ on $q_2 / \langle p_{\rm T} \rangle$ and collision system is because triangular harmonics in these small systems is predominantly fluctuation-driven and insensitive to the tip-tip or body-body orientation. As a result, $\langle \cos[6(\psi_3 - \psi_6)]\rangle_{\rm GE}$ dependence of $q_2 / \langle p_{\rm T} \rangle$ are primarily sensitive to the octupole deformations, such as the tetrahedral \(\alpha\)-cluster structure of $^{16}$O, rather than the overall event orientation. This complementary sensitivity allows $\langle \cos[4(\psi_2 - \psi_4)]\rangle_{\rm GE}$ and $\langle \cos[6(\psi_3 - \psi_6)]\rangle_{\rm GE}$ as a function of $q_2 / \langle p_{\rm T} \rangle$ to disentangle contributions from different underlying nuclear geometries in small collision systems.

\section{Summary}
\label{sec_summary}

In summary, we report the first study of symmetry plane correlations with an explicit focus on disentangling the role of intrinsic nuclear configurations in OO and Ne--Ne collisions at $\sqrt{s_{\rm NN}}=5.36$~TeV within the AMPT framework. Realistic ground-state configurations of $^{20}$Ne and $^{16}$O, obtained from NLEFT and PGCM, are employed. The important findings are:

\begin{enumerate}
    \item The correlator $\langle \cos[4(\psi_2 - \psi_4)]\rangle_{\rm GE}$ is strongly sensitive to the quadrupole-deformed (``pinball'') structure of $^{20}$Ne.
    
    \item The large quadrupole deformation of $^{20}$Ne leads to an enhanced $\langle \cos[4(\psi_2 - \psi_4)]\rangle_{\rm GE}$ in central Ne--Ne collisions relative to OO collisions.
    
    \item The correlator $\langle \cos[6(\psi_3 - \psi_6)]\rangle_{\rm GE}$ is larger in OO than in Ne--Ne collisions, reflecting its sensitivity to the intrinsic triangular symmetry of $^{16}$O.
    
    \item $\langle \cos[4(\psi_2 - \psi_4)]\rangle_{\rm GE}$ exhibits a clear dependence on the collision geometry (tip--tip versus body--body), whereas $\langle \cos[6(\psi_3 - \psi_6)]\rangle_{\rm GE}$ shows no significant dependence within uncertainties.
\end{enumerate}

These results establish symmetry plane correlations as a sensitive probe of initial-state geometry in small collision systems. In contrast to earlier studies emphasizing final-state collectivity, our findings provide direct evidence of sensitivity to intrinsic nuclear structure. The present results can be confronted with experimental data and other theoretical approaches, offering a novel avenue to constrain nuclear configurations through measurements of symmetry plane correlations.

\section*{Acknowledgment}
S.P. acknowledges the support from the Hungarian National Research, Development and Innovation Office (NKFIH) under the contract numbers NKFIH NKKP ADVANCED\_25-153456, 2025-1.1.5-NEMZ\_KI-2025-00005, 2024-1.2.5-TET-2024-00022, and the usage of Wigner Scientific Computing Laboratory (WSCLAB). R.S. acknowledges the DAE-DST, Government of India, funding under the mega-science project “Indian participation in the ALICE experiment at CERN” bearing Project No. SR/MF/PS-02/2021-IITI(E-37123). The authors gratefully acknowledge the MoU between IIT Indore and Wigner Research Centre for Physics (WRCP), Hungary, for the techno-scientific international cooperation.

\newpage


\begin{thebibliography}{99}

\bibitem{Heinz:2013th}
U.~Heinz and R.~Snellings,
Ann. Rev. Nucl. Part. Sci. \textbf{63}, 151 (2013).

\bibitem{Ollitrault:1992bk}
J.~Y.~Ollitrault,
Phys. Rev. D \textbf{46}, 229 (1992).

\bibitem{Voloshin:1994mz}
S.~Voloshin and Y.~Zhang,
Z. Phys. C \textbf{70}, 665 (1996).

\bibitem{Bhalerao:2011yg}
R.~S.~Bhalerao, M.~Luzum and J.~Y.~Ollitrault,
Phys. Rev. C \textbf{84}, 034910 (2011).

\bibitem{Prasad:2025ezg}
S.~Prasad, A.~M.~Kavumpadikkal Radhakrishnan, R.~Sahoo and N.~Mallick,
Phys. Lett. B \textbf{868}, 139753 (2025).

\bibitem{Prasad:2022zbr}
S.~Prasad, N.~Mallick, S.~Tripathy and R.~Sahoo,
Phys. Rev. D \textbf{107}, 074011 (2023).

\bibitem{Giacalone:2021udy}
G.~Giacalone, J.~Jia and C.~Zhang,
Phys. Rev. Lett. \textbf{127}, 242301 (2021).


\bibitem{Behera:2023nwj}
D.~Behera, S.~Prasad, N.~Mallick and R.~Sahoo,
Phys. Rev. D \textbf{108}, 054022 (2023).


\bibitem{ALICE:2022wpn}
S.~Acharya \textit{et al.} (ALICE Collaboration),
Eur. Phys. J. C \textbf{84}, 813 (2024).

\bibitem{Parkkila:2021tqq}
J.~E.~Parkkila, A.~Onnerstad and D.~J.~Kim,
Phys. Rev. C \textbf{104}, 054904 (2021).

\bibitem{Parkkila:2021yha}
J.~E.~Parkkila, A.~Onnerstad, S.~F.~Taghavi, C.~Mordasini, A.~Bilandzic, M.~Virta and D.~J.~Kim,
Phys. Lett. B \textbf{835}, 137485 (2022).

\bibitem{Prasad:2025yfj}
S.~Prasad, S.~Tripathy, B.~Sahoo and R.~Sahoo, Physics Reports {\bf 1181} (2026) 1–75.

\bibitem{Wu:2022exl}
X.~Wu, X.~Li, Z.~Tang, P.~Wang and W.~Zha,
Phys. Rev. Res. \textbf{4}, L042048 (2022).

\bibitem{E877:1996czs}
J.~Barrette \textit{et al.} (E877 Collaboration),
Phys. Rev. C \textbf{55}, 1420 (1997)
[erratum: Phys. Rev. C \textbf{56}, 2336 (1997)].

\bibitem{Bilandzic:2020csw}
A.~Bilandzic, M.~Lesch and S.~F.~Taghavi,
Phys. Rev. C \textbf{102}, 024910 (2020).

\bibitem{ALICE:2023wdn}
S.~Acharya \textit{et al.} (ALICE Collaboration),
Eur. Phys. J. C \textbf{83}, 576 (2023).

\bibitem{STAR:2005gfr}
J.~Adams \textit{et al.} (STAR Collaboration),
Nucl. Phys. A \textbf{757}, 102 (2005).

\bibitem{ATLAS:2014ndd}
G.~Aad \textit{et al.} (ATLAS Collaboration),
Phys. Rev. C \textbf{90}, 024905 (2014).


\bibitem{Brewer:2021kiv}
J.~Brewer, A.~Mazeliauskas and W.~van der Schee,
[arXiv:2103.01939 [hep-ph]].

\bibitem{Citron:2018lsq}
Z.~Citron, A.~Dainese, J.~F.~Grosse-Oetringhaus, and J.~M.~Jowett, \textit{et al.}
CERN Yellow Rep. Monogr. \textbf{7}, 1159 (2019).

\bibitem{GGamowbook}
G. Gamow,
Nature 130, 648 (1932). 

\bibitem{Wheeler:1937zza}
J.~A.~Wheeler,
Phys. Rev. \textbf{52}, 1083 (1937).

\bibitem{Lombardo:2025erk}
I.~Lombardo, L.~Redigolo, D.~Dell'Aquila, and M.~Vigilante, \textit{et al.}
Nucl. Phys. A \textbf{1060}, 123115 (2025)


\bibitem{ALICE:2025luc}
I.~J.~Abualrob \textit{et al.} (ALICE Collaboration),
arXiv:2509.06428.


\bibitem{Behera:2021zhi}
D.~Behera, N.~Mallick, S.~Tripathy, S.~Prasad, A.~N.~Mishra and R.~Sahoo,
Eur. Phys. J. A \textbf{58}, 175 (2022).

\bibitem{MenonKavumpadikkalRadhakrishnan:2025apq}
A.~Menon Kavumpadikkal Radhakrishnan, S.~Prasad, N.~Mallick, R.~Sahoo and G.~G.~Barnaf{\"o}ldi,
Phys. Lett. B \textbf{870}, 139941 (2025).

\bibitem{Prasad:2024ahm}
S.~Prasad, N.~Mallick, R.~Sahoo and G.~G.~Barnaf{\"o}ldi,
Phys. Lett. B \textbf{860}, 139145 (2025).

\bibitem{Li:2020vrg}
Y.~A.~Li, S.~Zhang and Y.~G.~Ma,
Phys. Rev. C \textbf{102}, 054907 (2020).

\bibitem{Bozek:2014cva}
P.~Bozek, W.~Broniowski, E.~Ruiz Arriola and M.~Rybczynski,
Phys. Rev. C \textbf{90}, 064902 (2014).

\bibitem{Broniowski:2013dia}
W.~Broniowski and E.~Ruiz Arriola,
Phys. Rev. Lett. \textbf{112}, 112501 (2014).


\bibitem{Behera:2023oxe}
D.~Behera, S.~Deb, C.~R.~Singh and R.~Sahoo,
Phys. Rev. C \textbf{109}, 014902 (2024).

\bibitem{Ding:2023ibq}
C.~Ding, L.~G.~Pang, S.~Zhang and Y.~G.~Ma,
Chin. Phys. C \textbf{47}, 024105 (2023).

\bibitem{Wang:2021ghq}
Y.~Z.~Wang, S.~Zhang and Y.~G.~Ma,
Phys. Lett. B \textbf{831}, 137198 (2022).

\bibitem{Rybczynski:2017nrx}
M.~Rybczy\'nski, M.~Piotrowska and W.~Broniowski,
Phys. Rev. C \textbf{97}, 034912 (2018).

\bibitem{Svetlichnyi:2023nim}
A.~Svetlichnyi, S.~Savenkov, R.~Nepeivoda and I.~Pshenichnov,
MDPI Physics \textbf{5}, 381 (2023).


\bibitem{Giacalone:2024ixe}
G.~Giacalone, W.~Zhao, B.~Bally, and S.~Shen, \textit{et al.}
Phys. Rev. Lett. \textbf{134}, 082301 (2025).


\bibitem{Zhang:2024vkh}
C.~Zhang, J.~Chen, G.~Giacalone, S.~Huang, J.~Jia and Y.~G.~Ma,
Phys. Lett. B \textbf{862}, 139322 (2025).

\bibitem{R:2024eni}
A.~Menon Kavumpadikkal Radhakrishnan, S.~Prasad, N.~Mallick and R.~Sahoo,
Eur. Phys. J. A \textbf{61}, 134 (2025).

\bibitem{ATLAS:2025nnt}
G.~Aad \textit{et al.} (ATLAS Collaboration),
arXiv:2509.05171.


\bibitem{CMS:2025tga}
A.~Hayrapetyan \textit{et al.} (CMS Collaboration),
arXiv:2510.02580.


\bibitem{Tripathy:2025npe}
S.~Tripathy, S.~Prasad and R.~Sahoo,
Phys. Rev. D \textbf{112}, 114012 (2025).

\bibitem{Frosini:2021fjf}
M.~Frosini, T.~Duguet, J.~P.~Ebran and V.~Som{\`a},
Eur. Phys. J. A \textbf{58}, 62 (2022).

\bibitem{Frosini:2021sxj}
M.~Frosini, T.~Duguet, J.~P.~Ebran, B.~Bally, T.~Mongelli, T.~R.~Rodr{\'\i}guez, R.~Roth and V.~Som{\`a},
Eur. Phys. J. A \textbf{58}, 63 (2022).


\bibitem{Lahde:2019npb}
T.~A.~L{\"a}hde and U.~G.~Mei{\ss}ner,
Lect. Notes Phys. \textbf{957}, 1 (2019).

\bibitem{Lee:2020meg}
D.~Lee,
Front. in Phys. \textbf{8}, 174 (2020).

\bibitem{Loizides:2025ule}
C.~Loizides,
Phys. Rev. C \textbf{113}, 1 (2026).

\bibitem{Giacalone:2024luz}
G.~Giacalone, B.~Bally, G.~Nijs, and S.~Shen, \textit{et al.}
Phys. Rev. Lett. \textbf{135}, 012302 (2025).



\bibitem{Lin:2004en}
Z.~W.~Lin, C.~M.~Ko, B.~A.~Li, B.~Zhang and S.~Pal,
Phys. Rev. C \textbf{72}, 064901 (2005).

\bibitem{Zhang:1999bd}
B.~Zhang, C.~M.~Ko, B.~A.~Li and Z.~w.~Lin,
Phys. Rev. C \textbf{61}, 067901 (2000)

\bibitem{Wang:1991hta}
X.~N.~Wang and M.~Gyulassy,
Phys. Rev. D \textbf{44}, 3501 (1991).

\bibitem{Zhang:1997ej}
B.~Zhang,
Comput. Phys. Commun. \textbf{109}, 193 (1998).

\bibitem{Andersson:1983ia}
B.~Andersson, G.~Gustafson, G.~Ingelman and T.~Sjostrand,
Phys. Rept. \textbf{97}, 31 (1983).

\bibitem{Greco:2003xt}
V.~Greco, C.~M.~Ko and P.~Levai,
Phys. Rev. Lett. \textbf{90}, 202302 (2003).

\bibitem{Li:1995pra}
B.~A.~Li and C.~M.~Ko,
Phys. Rev. C \textbf{52}, 2037 (1995).

\bibitem{Li:2001xh}
B.~Li, A.~T.~Sustich, B.~Zhang and C.~M.~Ko,
Int. J. Mod. Phys. E \textbf{10}, 267 (2001).


\bibitem{STAR:2002hbo}
C.~Adler \textit{et al.} (STAR Collaboration),
Phys. Rev. C \textbf{66}, 034904 (2002).

\bibitem{ATLAS:2015qwl}
G.~Aad \textit{et al.} (ATLAS Collaboration),
Phys. Rev. C \textbf{92}, 034903 (2015).

\bibitem{Teaney:2013dta}
D.~Teaney and L.~Yan,
Phys. Rev. C \textbf{90}, 024902 (2014).

\bibitem{Gardim:2011xv}
F.~G.~Gardim, F.~Grassi, M.~Luzum and J.~Y.~Ollitrault,
Phys. Rev. C \textbf{85}, 024908 (2012).

\end{thebibliography}
\end{document}